\documentclass[aps, a4paper, amsfonts, amssymb, amsmath, reprint, showkeys, nofootinbib, twoside, showkeys, normalem, showpacs, superscriptaddress, epsfig, floatfix, normalem reprint, twoside, twocolumn, preprintnumbers]{revtex4-2}

\usepackage[utf8]{inputenc}
\usepackage[english]{babel}
\usepackage{graphicx}
\usepackage{dcolumn}
\usepackage{bm}
\usepackage{graphicx,color,xcolor}
\usepackage{physics}
\usepackage{comment}
\usepackage{ulem}
\usepackage{mathtools}
\usepackage{mathptmx}
\usepackage{amsmath,amssymb}
\usepackage{mathrsfs}
\graphicspath{ {./Figures/} }
\usepackage{epsfig}
\usepackage{braket}
\usepackage{amsmath}
\usepackage{hyperref}
 \hypersetup{
     colorlinks=true,
     linkcolor=blue,
     filecolor=blue,
     citecolor = blue,      
     urlcolor=blue}
     
\usepackage{graphicx,color,xcolor,colortbl}

\definecolor{LightCyan}{rgb}{0.88,1,1}

\DeclareUnicodeCharacter{2212}{Here!!!}

\begin{document}
\title{Ultrafast switching of topological invariants by light-driven strain}

\author{Tae Gwan Park}
\affiliation{Department of Physics, Korea Advanced Institute of Science and Technology (KAIST), Daejeon 34141, Republic of Korea}

\author{Seungil Baek}
\affiliation{Department of Physics, Korea Advanced Institute of Science and Technology (KAIST), Daejeon 34141, Republic of Korea}

\author{Junho Park}
\affiliation{Department of Physics, Korea Advanced Institute of Science and Technology (KAIST), Daejeon 34141, Republic of Korea}

\author{Eui-Cheol Shin}
\affiliation{Department of Physics, Korea Advanced Institute of Science and Technology (KAIST), Daejeon 34141, Republic of Korea}

\author{Hong Ryeol Na}
\affiliation{Department of Physics and Astronomy, Sejong University, Seoul 05006, Republic of Korea}

\author{Eon-Taek Oh}
\affiliation{Department of Physics, Korea Advanced Institute of Science and Technology (KAIST), Daejeon 34141, Republic of Korea}

\author{Seung-Hyun Chun}
\affiliation{Department of Physics and Astronomy, Sejong University, Seoul 05006, Republic of Korea}

\author{Yong-Hyun Kim}
\email{yong.hyun.kim@kaist.ac.kr}
\affiliation{Department of Physics, Korea Advanced Institute of Science and Technology (KAIST), Daejeon 34141, Republic of Korea}

\author{Sunghun Lee}
\email{kshlee@sejong.ac.kr}
\affiliation{Department of Physics and Astronomy, Sejong University, Seoul 05006, Republic of Korea}

\author{Fabian Rotermund}
\email{rotermund@kaist.ac.kr}
\affiliation{Department of Physics, Korea Advanced Institute of Science and Technology (KAIST), Daejeon 34141, Republic of Korea}

\date{\today}

\begin{abstract}
Reversible control of the topological invariants from nontrivial to trivial states has fundamental implications for quantum information processors and spintronics, by realizing of an on/off switch for robust and dissipationless spin-current. Although mechanical strain has typically advantageous for such control of topological invariants, it is often accompanied by in-plane fractures and is not suited for high-speed, time-dependent operations. Here, we use ultrafast optical and THz spectroscopy to investigate topological phase transitions by light-driven strain in Bi$_2$Se$_3$, a material that requires substantial strain for $\mathrm{Z}_2$ switching. We show that Bi$_2$Se$_3$ experiences ultrafast switching from being a topological insulator with spin-momentum-locked surfaces, to hybridized states and normal insulating phases at ambient conditions. Light-induced strong out-of-plane strain can suppress the surface-bulk coupling, enabling differentiation of surface and bulk conductance at room temperature, far above the Debye temperature. We illustrate various time-dependent sequences of transient hybridization, as well as the switching operation of topological invariants by adjusting the photoexcitation intensity. The abrupt alterations in both surface and bulk transport near the transition point allow for coherent conductance modulation at hyper-sound frequencies. Our findings regarding light-triggered ultrafast switching of topological invariants pave the way for high-speed topological switching and its associated applications. 
\end{abstract}

\maketitle

\setcounter{figure}{0}
\renewcommand{\figurename}{Fig.}
\renewcommand{\thefigure}{\arabic{figure}}

Topological surface states (TSSs) are unique quantum states on topologically nontrivial insulators, induced by spin-orbit coupling~\cite{1,2}. These symmetry-protected helical TSSs provide robust and dissipationless transport channels, making them ideal for applications in spintronics and quantum computing~\cite{3,4}. Controlling the topological invariants allows for the implementation of a topological on/off switching device as a transistor. This constitutes a fundamental building block for future applications and the emergence of topology from trivial matter~\cite{5}. Although traditional methods like chemical substitution have successfully demonstrated topological phase transitions via ionic interactions modification~\cite{6,7,8,9}, additional features such as reversibility and time-dependent operation are required for effective switching devices. Mechanical strain can induce topological phase transitions and can potentially serve this purpose~\cite{10,11}, but it restricts high-speed and time-dependent operations. Alternatively, light has been identified as a promising alternative to control topology on an ultrafast timescale using light-driven phonons~\cite{12,13,14,15} and photocurrents~\cite{16,17}, as recently shown in Dirac and Weyl semimetals. However, this transition has not yet been observed in topological insulators (TIs) with a large bulk bandgap ($\mathrm{E}_{g}$), because they obviously require substantial strain. Furthermore, transport dynamics during topological phase transitions, which are one of the pivotal characteristics for topological switching applications, remain unexplored.

Bi$_2$Se$_3$ is a representative TI, which exhibits nontrivial $\mathrm{Z}_2$ order and has a layered rhombohedral structure with a quintuple layer (QL) unit. The large $\mathrm{E}_{g}$ of about 0.3 eV in bulk states (BS) of Bi$_2$Se$_3$ ensures room temperature topological applications~\cite{3,4,18}. However, topological phase transitions in Bi$_2$Se$_3$ require a large strain above 5\% along out-of-plane direction~\cite{19}, which is impossible to achieve with mechanical strain due to unavoidable in-plane fractures, formed around 1\% longitudinal strain~\cite{20}. Previous studies of longitudinal straining and lattice vibrations by ultrafast coherent light motivate us to examine light-driven ultrafast $\mathrm{Z}_2$ switching in Bi$_2$Se$_3$~\cite{21,22,23}. Here, we utilize ultrafast light pulses to selectively apply longitudinal strain as a means to induce the topological phase transition in Bi$_2$Se$_3$. We find that light can induce a strain of about 7\%, sufficient to alter topological invariants. The layered structure with van der Waals (vdW) bonding in Bi$_2$Se$_3$ along the c-axis ensures durability even under such high strain. The lattice strain enables selective probing of TSS and BS charge transport at room temperature by suppressing the surface-bulk coupling, even far above the Debye temperature ($\sim$180 K)~\cite{24}, where phonon scattering predominates. During light-driven topological switching, the transport lifetime in TSS suddenly reduces to approximately 50\% due to light-induced hybridization. Simultaneously, the bulk conductance increases considerably during band parity inversion. We also demonstrate temporal sequences of light-driven $\mathrm{Z}_2$ switching, transient hybridization, and coherent modulation of transport in ambient conditions.

\begin{figure*}[t]%
\centering
\includegraphics[width=0.9\textwidth]{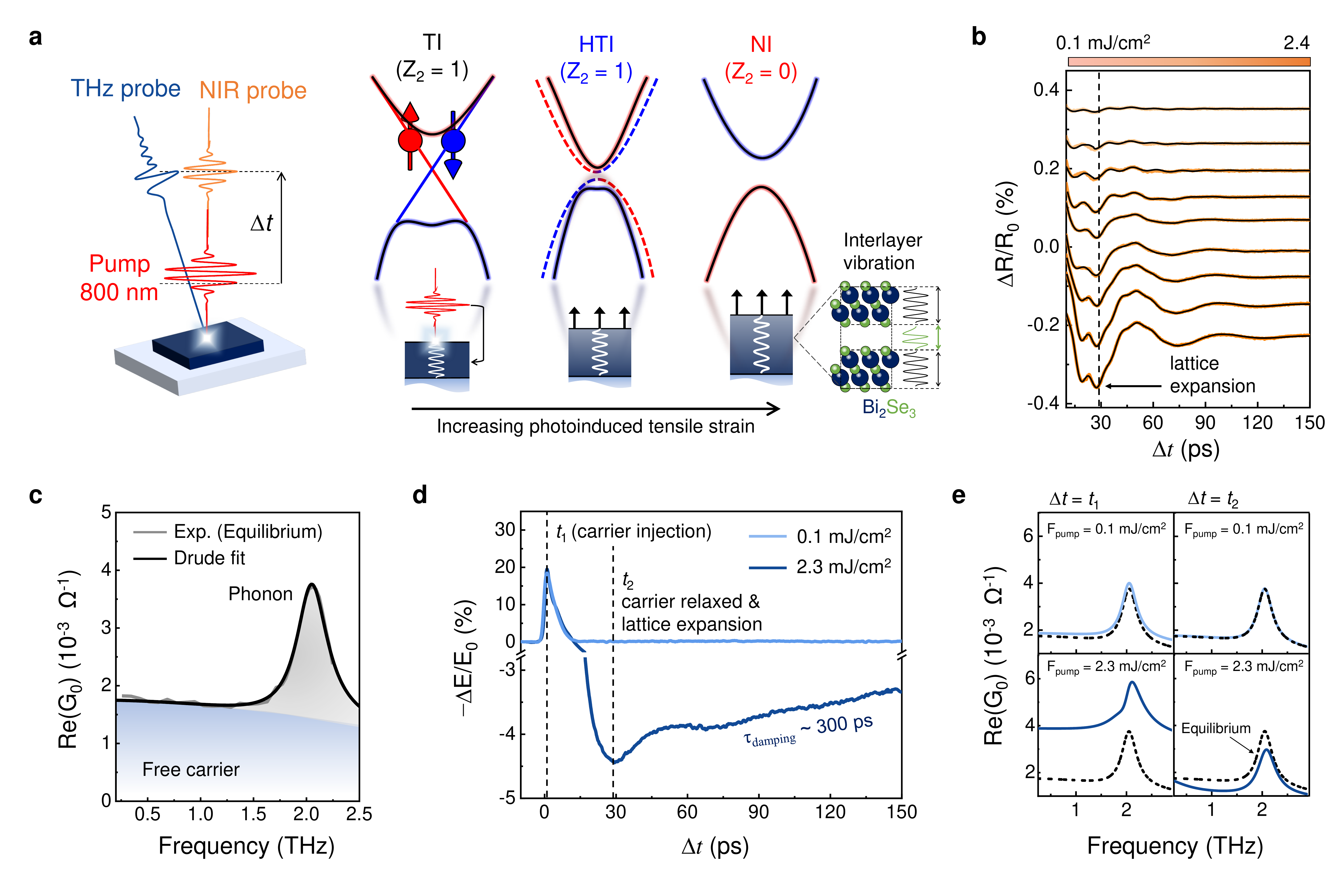}
\caption{
\textbf{Experimental scheme and light-driven strain waves and photoconductance dynamics in Bi$_\textbf{2}$Se$_\textbf{3}$.}
\textbf{a,} Schematics of photoinduced topological phase transitions and their probing by time-resolved ultrafast optical and THz spectroscopy. The pump pulses generate the photocarrier and subsequent tensile strain, which induces the expansion of Bi$_2$Se$_3$ thin film, consisting of the intralayer thickness (black spring) and interlayer distance (green spring). The strong out-of-plane straining in Bi$_2$Se$_3$ induces to the inversion of conduction/valence parity near the Gamma point, leading to a topological phase transition from topological insulator (TI) to hybridized TI (HTI) and normal insulator (NI). 
\textbf{b,} $\mathrm{F}_\mathrm{pump}$-dependent oscillatory signal measured in reflection changes at NIR wavelength. The black lines are the fit results of the experimental data with damped oscillations.
\textbf{c,} Real part of THz conductance of 22 QL Bi$_2$Se$_3$ at equilibrium obtained by THz time-domain spectroscopy. The black line represents the Drude fit. 
\textbf{d}, Temporal evolution in photoexcited dynamics, measured with conductance changes in THz probe ($- \Delta$E/$\mathrm{E}_0$) with $\mathrm{F}_\mathrm{pump}$ = 0.1 mJ/$\mathrm{cm}^2$ and 2.3 mJ/$\mathrm{cm}^2$.
\textbf{e,} Real part of THz conductance by 0.1 mJ/$\mathrm{cm}^2$ and 2.3 mJ/$\mathrm{cm}^2$ photoexcitation at selected time delay ($t_{1}$ and $t_{2}$) with equilibrium THz conductance (dashed black curves). The maximum in photoconductance at $t_{1}$ implies the required time for relaxation of excited bulk carriers from the higher to the first conduction band. The $t_{2}$ indicates the time of maximum expansion.
}
\label{fig1}
\end{figure*}

Our experimental scheme involves a sequence of three pulses: an optical pump, a near-infrared (NIR) probe, and a terahertz (THz) probe, as depicted in Fig.~\ref{fig1}a. The optical pump with 1.5-eV photon energy produces photoinduced stress and strain within the Bi$_2$Se$_3$ film by means of excited carriers. The resulting strain and simultaneous transport characteristics are monitored using NIR and THz probe pulses. The atomic displacement in Bi$_2$Se$_3$ by photoinduced stress is perturbed with quasi-spherical symmetry due to its lateral isotropy. Given experimental conditions—with large beam spots on the sample (denoted A, at the scale of micrometers and millimeters for NIR and THz measurements, respectively) compared to the optical penetration depth ($\xi\sim$20 nm)~\cite{23} for pump pulses ($A \gg \xi$)—shear and quasi-shear stresses are effectively nullified. This is because the transverse displacement is orthogonal to the spherical symmetry~\cite{25}. Thus, the photoinduced strain waves propagate as longitudinal plane waves (see Fig.~\ref{figS1}). The longitudinal strain ($\eta_{33}$) confined within the film forms standing waves and acts as an effective tensile strain when the film thickness ($d$) is reduced to the nanoscale~\cite{23}. This feature makes it notably different from mechanical strain, which inevitably results in in-plane elongation. In a layered structure, both the intra-QL thickness and the inter-QL distance are influenced by $\eta_{33}$, as shown by the black and green springs in Fig. 1a. Due to the weak vdW bonding, density functional theory (DFT) calculations reveal that the inter-QL distance changes easier than the intra-QL thickness (see Fig.~\ref{figS2}). This aspect makes the crystal resilient against large expansion in the stacking direction. This substantial inter-QL distance can expand the Coulomb gap and reduce the strength of the spin-orbit interaction~\cite{19}, leading to the inversion of conduction/valence band parity, and thus, to topological phase transitions and hybridized states as illustrated in Fig.~\ref{fig1}a.

The changes in the real part of the refractive index ($\tilde{n}$) caused by deformation are recorded by NIR probe pulses. Since the sensitivity for measuring strain is mostly pronounced near the bandgap (i.e., $\mathrm{A}_\mathrm{osc}\sim d\tilde{n}/d\omega\cdot\delta$$E_{g})$~\cite{26}, the probe energy is selected as 0.92 eV (1350 nm in wavelength) which is close to the bandgap between the first bulk conduction and the second valence band~\cite{27} (Fig.~\ref{figS3}). As illustrated in Fig.~\ref{fig1}b, the $\Delta$R/R$_0$ signal shows the transient strain $\eta_{33}$ (i.e., expansion reduces the probe reflection), which is equivalent to out-of-plane interlayer vibrations~\cite{21,23}. Simultaneously, THz pulses allow us to monitor electrical conductance under photoinduced strain. Figure~\ref{fig1}b displays the real part of THz sheet conductance ($\mathrm{G}_0$) of 22 QL Bi$_2$Se$_3$, incorporating Drude-Lorentzian terms from free carrier (Lorentzian center, $\omega_L=0$) and optical phonon in the bulk ($\omega_L\sim2$ THz)~\cite{7}. In the Drude model, the bases for TSS are the 2D spectral weight ($D_\mathrm{TSS}=\omega_p^2d/4\pi^2$) and the scattering rate ($\gamma_\mathrm{TSS}= 1/\tau_\mathrm{TSS}$), where $\omega_p$ and $\tau_\mathrm{TSS}$ are the plasma frequency and transport lifetime, respectively. The $D_\mathrm{TSS}$ value derived from the fit is 138 $\mathrm{THz}^2\cdot$QL, corresponding to a sheet carrier density ($n_{2D}=4\pi^2m^*\epsilon_0D_\mathrm{TSS}/e^2$) of 1.78 × 10$^{13}$ $\mathrm{cm}^{-2}$ and a Fermi level ($E_{F}=2\pi \hbar D_\mathrm{TSS}/15e^2$) of 340 meV~\cite{6,27,28}, where $m^*=0.15m_0$ is the electron effective mass~\cite{29}. The measured Fermi level is slightly above the bulk bandgap, suggesting that the THz response primarily arises from TSS. Given that our experiments were conducted at room temperature, above the Debye temperature, phonon-mediated surface-bulk coupling results in a flat spectrum with $\gamma_\mathrm{TSS}\sim 3.9$ THz. The calculated mobility ($\mu=e\tau_\mathrm{TSS}/m^*$) is 254 $\mathrm{cm}^2$/V$\cdot$s, indicating that transport properties are mainly influenced by phonon scattering. It is worth noting that separating the TSS response from the phonon-mediated surface-bulk coupling remains a major challenge at room temperature~\cite{30,31}, which is crucial for operating topological devices in an ambient environment.

\begin{figure*}[t]%
\centering
\includegraphics[width=0.9\textwidth]{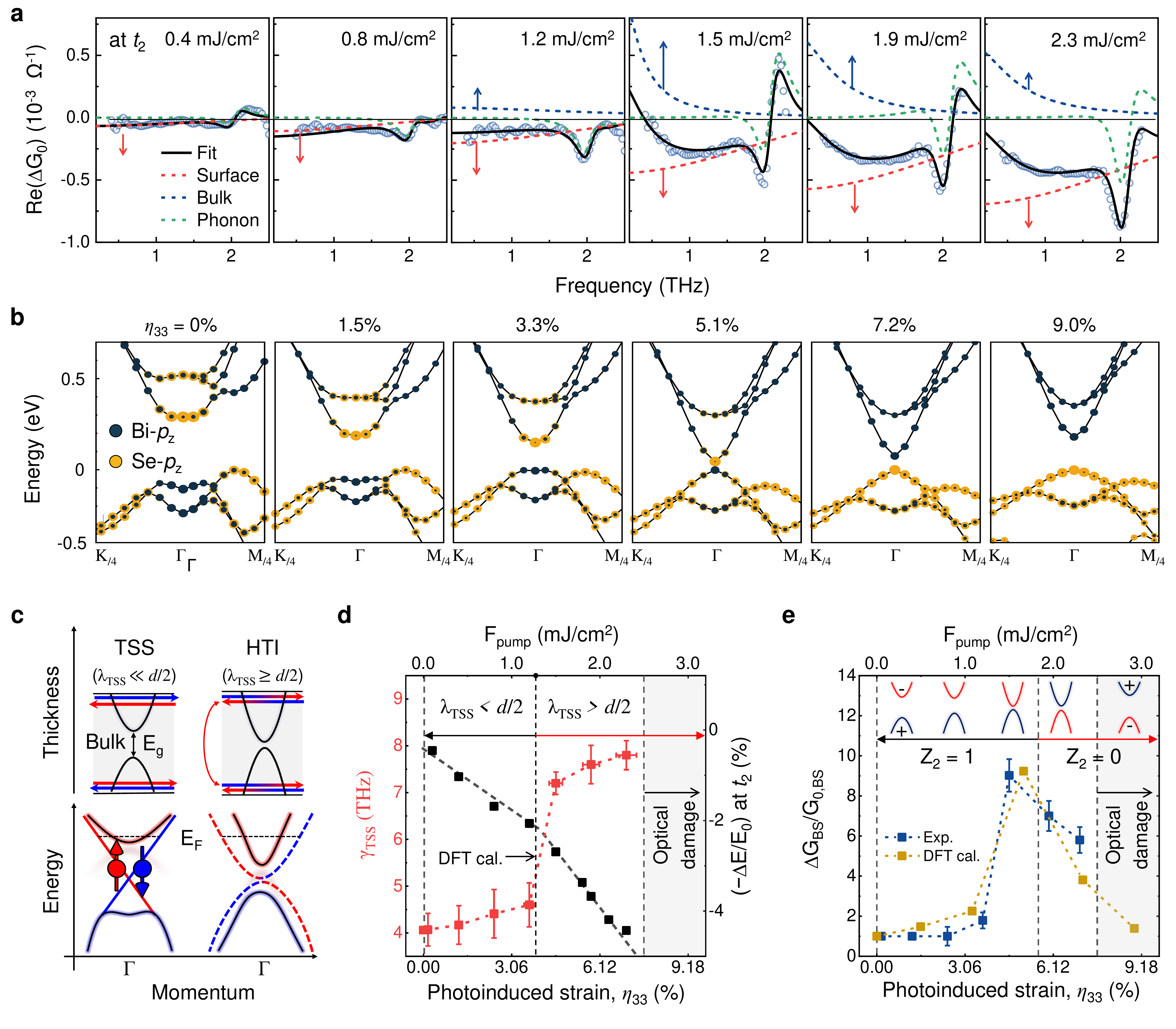}
\caption{
\textbf{Manipulation of TSS and BS transport across topological phase transition by light-driven tensile strain.}
\textbf{a,} Differential THz conductance ($\Delta\mathrm{G}_0$) of Bi$_2$Se$_3$ at various $\mathrm{F}_\mathrm{pump}$ recorded at $t_{2}$, where the lattice maximally expands. 
\textbf{b,} Orbital-decomposed band structure of Bi$_2$Se$_3$ bulk from DFT calculation according to tensile strain. The orbital contributions from Bi-$p_z$ and Se-$p_z$ are marked with blue and orange color circles, where the size of the circle represents the magnitude of the contribution. Here, the K/4 and M/4 are the quarter of K$-\Gamma$ and $\Gamma -$M. 
\textbf{c,} Schematics of TSS hybridization and suppression of surface-bulk coupling by tensile strain.
\textbf{d}, $\mathrm{F}_\mathrm{pump}$- and photoinduced strain-dependent scattering rate in TSS and $- \Delta$E/$\mathrm{E}_0$ values at $t_{2}$. The photoinduced strain is obtained by dividing the calculated $\sigma_{33}$ by the bulk modulus ($C_{33}$). The TSS scattering rate ($\gamma_\mathrm{TSS}$, inverse of transport lifetime in TSS), obtained by the fit results from (\textbf{a}) suddenly increases for about 1.3 mJ/$\mathrm{cm}^2$, corresponding to about 4\% strain, which results in the slope changes in $- \Delta$E/$\mathrm{E}_0$ obtained at $t_{2}$ as indicated by the vertical dashed line. The vertical orange line denote the transition point of hybridized topological insulator (HTI) from DFT prediction. The transition point of NI is obtained from the (\textbf{b}). The optical damage is observed above $\mathrm{F}_\mathrm{pump}=$ 2.5 mJ/$\mathrm{cm}^2$.
\textbf{e,} Bulk conductance changes according to $\mathrm{F}_\mathrm{pump}$ and strain adopted from the data in (\textbf{a}) and DFT calculations. The experimental $\Delta \mathrm{G}_\mathrm{BS}$ is obtained from the simple relation of $\Delta \mathrm{G}_\mathrm{BS} \sim 1/\Delta\gamma_\mathrm{BS}$. The bulk conductance shows a significant increase across the topological phase transition at the same level of strain as indicated by the red vertical dashed line. The error bars in (\textbf{d}) and (\textbf{e}) represent the standard deviation uncertainties of the fitting results.
}
\label{fig2}
\end{figure*}

Figures~\ref{fig1}d and ~\ref{fig1}e show the dynamics of photoconductance for pump fluence ($\mathrm{F}_\mathrm{pump}$) of 0.1 and 2.3 mJ/$\mathrm{cm}^2$ with THz spectra at selected times $t_{1}$ (carrier injection) and $t_{2}$ (carrier relaxation and lattice expansion). Given that the measured $\mathrm{G}_0$ is inversely proportional to THz transmission, Fig. 1d depicts the THz conductance deviated from equilibrium, as derived from frequency integration (i.e., $-\Delta \mathrm{E}/\mathrm{E}_0 \propto\Delta\mathrm{G}_0$). Therefore, an initial rise in $-\Delta \mathrm{E}/\mathrm{E}_0$ at $t_{1}$ corresponds to an increase in $\mathrm{G}_0$ due to photocarrier injection in BS ~\cite{27,28}. The excited carrier relaxes quickly within 10 ps with the bulk relaxation time ($\tau_\mathrm{bulk}=$ 2 ps)~\cite{32,33}, and the unrelaxed carriers at $t_{2}$, estimated from the Drude fit, are negligible, constituting just 0.12\% of excited carriers ($n_{ex}$), as shown in Fig.~\ref{figS4}. Following carrier relaxation, a significant reduction in $\Delta\mathrm{G}_0$ with $\mathrm{F}_\mathrm{pump}$ = 2.3 mJ/$\mathrm{cm}^2$ is observed, accompanied by subsequent coherent modulation, which correlates with the period in NIR probing (Fig.~\ref{fig1}b). This negative $\Delta\mathrm{G}_0$ is noticeable above 0.4 mJ/$\mathrm{cm}^2$ (see Fig.~\ref{figS5}a). Given that long-lived carriers in TSS increase $\Delta\mathrm{G}_0$~\cite{33}, the observed negative $\Delta\mathrm{G}_0$ is reproduced by the increased $\gamma_\mathrm{TSS}$. Note that the negative $\Delta\mathrm{G}_0$ at $t_{2}$ coincides with the time required for maximum lattice expansion as observed in $\Delta$R/$\mathrm{R}_0$, similar to a recent report on ultrafast X-ray experiments~\cite{22}. In addition, the damping time ($\tau_\mathrm{damping}$) of negative photoconductance at $\sim$300 ps (refer to Fig.~\ref{fig1}d and S5b) matches the strain damping time ($\sim$300–500 ps) observed in ultrafast X-ray experiments, whereas thermal relaxation occurs on a much longer timescale of a few nanoseconds~\cite{22,28}. As a result, the observed decrease in THz conductance is primarily influenced by photoinduced tensile strain.

\begin{figure*}[t]%
\centering
\includegraphics[width=0.8\textwidth]{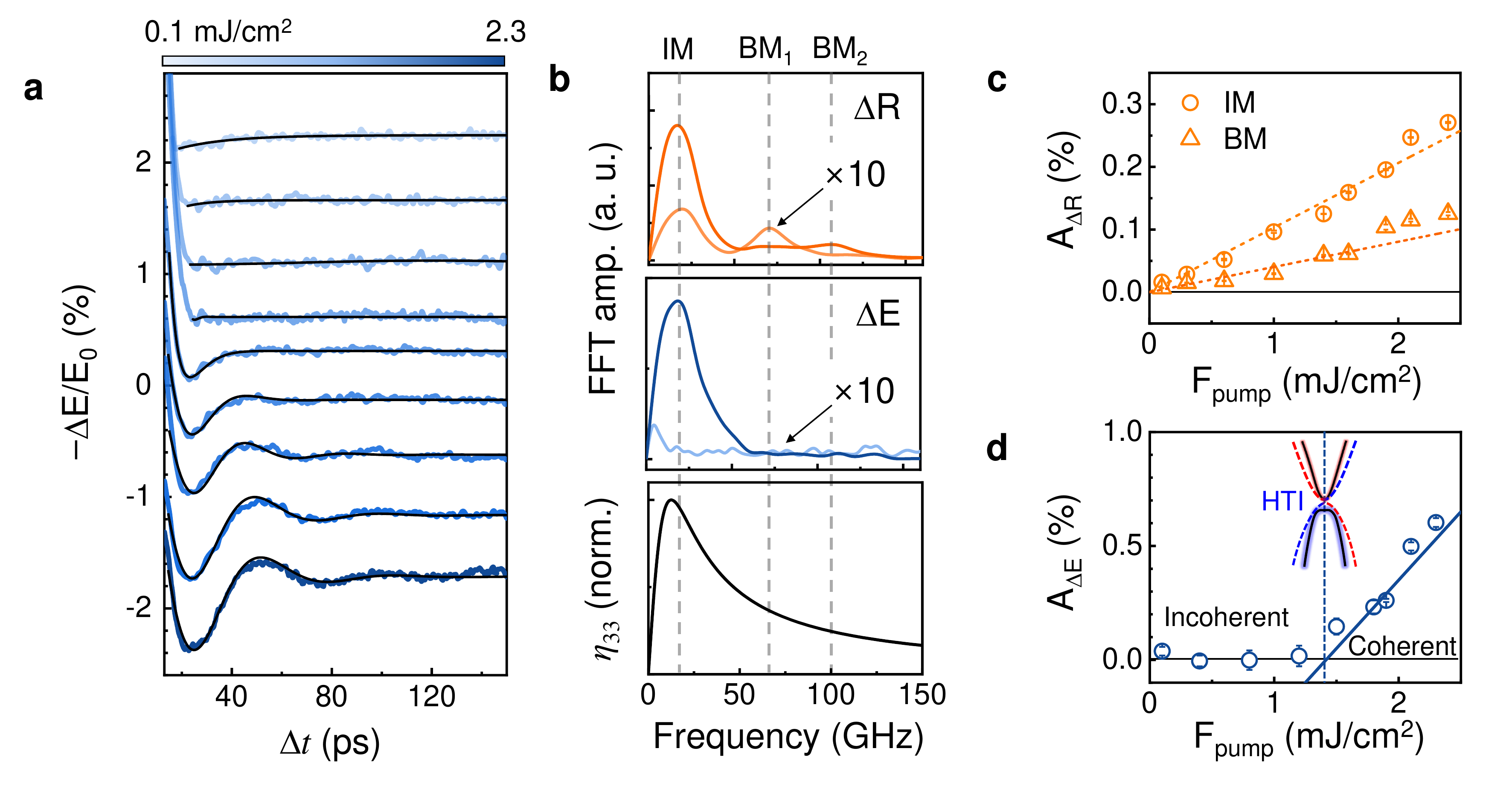}
\caption{
\textbf{Coherent modulation of photoconductance in Bi$_2$Se$_3$ by interlayer vibration.}
\textbf{a,} $\mathrm{F}_\mathrm{pump}$-dependent oscillatory signal measured with THz probe ($- \Delta$E/$\mathrm{E}_0$). The black lines are the fit results of the experimental data with damped oscillation.
\textbf{b,} Fast Fourier transform spectra of oscillatory signal in NIR probe from Fig.~\ref{fig1}b and THz probe from (\textbf{a}). In NIR probe, three oscillation components can be assigned by interlayer vibrational modes including interfacial mode (IM, 20 GHz), 1st breathing mode ($\mathrm{BM}_1$, 70 GHz), and 2nd breathing mode ($\mathrm{BM}_2$, 100 GHz). While in THz probe, only IM involves the modulation of transport characteristics in Bi$_2$Se$_3$. The bottom panel indicates the simulated spectra of photoinduced strain waves ($\eta_{33}$). The strain spectra explain the amplitude difference between observed vibrational modes. 
\textbf{c,} Measured oscillation amplitude in NIR probe ($\mathrm{A}_{\Delta \mathrm{R}}$) as a function of $\mathrm{F}_\mathrm{pump}$. The amplitude BM is produced by adding $\mathrm{BM}_1$ and $\mathrm{BM}_2$. The dashed line corresponds to the linear fit of the experimental data. The observed linear relation of $\mathrm{F}_\mathrm{pump}$ and $\mathrm{A}_{\Delta \mathrm{R}}$ gives a reliability of the linearity of strain amplitude upon the $\mathrm{F}_\mathrm{pump}$.
\textbf{d}, Measured oscillation Amplitude in THz probe ($\mathrm{A}_{\Delta \mathrm{E}}$) as a function of $\mathrm{F}_\mathrm{pump}$. The modulation with IM frequency is invisible at low $\mathrm{F}_\mathrm{pump}$ (incoherent modulation). Above the threshold of $\mathrm{F}_\mathrm{pump} \sim $ 1.3 mJ/$\mathrm{cm}^2$, coherent modulation is observed and shows a linear dependence of $\mathrm{F}_\mathrm{pump}$. The threshold fluence is consistent with $\mathrm{F}_\mathrm{pump}$ for HTI as indicated by the vertical blue line. This means that the coherent modulation requires substantial changes in transport characteristics. The error bars in (\textbf{c}) and (\textbf{d}) represent the standard deviation uncertainties of the measured data.
}
\label{fig3}
\end{figure*}

The temporal separation of lattice dynamics from ultrafast carrier dynamics allows us to study transport properties under strain. Figure~\ref{fig2}a presents the $\mathrm{F}_\mathrm{pump}$-dependent $\Delta \mathrm{G}_0$ spectra at $t_{2}$ by varying $\mathrm{F}_\mathrm{pump}$ with Drude fits. During the fitting process, $\Delta D_\mathrm{TSS}$ maintains a linear dependence on $\mathrm{F}_\mathrm{pump}$, while the unrelaxed carriers are at a negligible level of about 0.12\% of $n_{ex}$. Below 0.8 mJ/$\mathrm{cm}^2$, a slight decrease in conductance is fitted with an increase in $\gamma_\mathrm{TSS}$ as well as broadening and shift of optical phonon frequency~\cite{27}. At 1.2 mJ/$\mathrm{cm}^2$, the lower-energy tail in the spectra becomes flat, indicating a competing channel with increased conductivity, which becomes markedly visible at higher $\mathrm{F}_\mathrm{pump}$. The observed bipolar behavior in $\Delta \mathrm{G}_0$ can be understood by considering the strain effects at both TSS and BS. Qualitatively, surface conduction vanishes during the topological phase transition, while bulk conduction turns metallic due to gap closing. To analyze this quantitatively, we compared experimental results with DFT calculations. Figure~\ref{fig2}b displays the DFT-calculated electronic band structure with increasing tensile strain along the out-of-plane direction, which shows the $\mathrm{E}_{g}$ closing and subsequent $\mathrm{Z}_2$ switching. For TSS under varying $\mathrm{E}_{g}$, the $\gamma_\mathrm{TSS}$ should increase due to a finite size effect~\cite{6,34,35}. Since the penetration depth of TSS wave function ($\lambda_\mathrm{TSS}\sim \hbar v_F/\mathrm{E}_g$) into bulk Bi$_2$Se$_3$ is about 2.5 nm for $\mathrm{E}_{g}$ = 0.3 eV~\cite{6,35}, TSSs in 22 QL samples (thickness, $d \approx$ 22 nm) are initially isolated at the top and bottom surfaces with opposite spin chirality. However, the photoinduced tensile strain causes a reduction of $\mathrm{E}_{g}$, leading to an increase in $\lambda_\mathrm{TSS}$. When $\lambda_\mathrm{TSS}$ is close to half the $d$, the top and bottom TSS overlap, leading to hybridization, as shown in Fig.~\ref{fig2}c. This hybridized TI (HTI) expands the phase space for carrier scattering, including 180° backscattering~\cite{6,7,36}.

Figure~\ref{fig2}d illustrates the transport characteristics in TSS as a function of $\mathrm{F}_\mathrm{pump}$ and $\eta_{33}$. These characteristics were derived from the Drude analysis depicted in Fig.~\ref{fig2}a. To compare with DFT calculations, we convert $\mathrm{F}_\mathrm{pump}$ into strain by simulating the photoinduced stress ($\sigma_{33}$), taking into account both electronic stress ($\sigma_{e}$) and thermal stress ($\sigma_{t}$)~\cite{37}, as shown in Fig.~\ref{figS6}. The negative sign of $\sigma_{33}$ indicates expansion. The electronic stress ($\sigma_{e}$) dominates over thermal expansion ($\sigma_{t}$) by a factor of 5, consistent with previous observations in $\mathrm{Bi}_2\mathrm{Te}_3$~\cite{38}. The total stress induced by laser pulses, calculated to be ~4 GPa, and the corresponding strain of ~7\% can be derived by considering the bulk modulus~\cite{19}. As illustrated in Fig.~\ref{fig2}d, the $\gamma_\mathrm{TSS}$  slightly increases, likely due to lattice heating and expansion. However, it undergoes a sudden surge beyond a 4\% strain. In the DFT calculation with longitudinal strain at 4\%, $\mathrm{E}_{g}$ is reduced by 5 times to $\sim$70 meV (Figs.~\ref{figS7} and ~\ref{figS8}), which subsequently extends the penetration depth of $\lambda_\mathrm{TSS}$ to $\sim$13 nm. This extended $\lambda_\mathrm{TSS}$, close to half of the $d$, is sufficient to trigger hybridization. Consequently, we observe a rapid drop in the transport lifetime in TSS to approximately 50\% in the HTI state. This means that the surface electrons are freeze-out. The phase succeeding the HTI is anticipated to be a normal insulator (NI) with a topologically trivial state, which could be further identified by observing the strain-dependent BS conductance.

The stain-dependent transport of the BS results in a topological metal phase, as the band parities converge toward inversion. Consequently, the additional channel with increased $\Delta \mathrm{G}_0$ corresponds to the bulk conductance influenced by strain. Figure~\ref{fig2}e illustrates the changes in the measured $\Delta \mathrm{G}_\mathrm{BS}$, compared with DFT calculations. The theoretical DC conductivity in BS, derived from fully anisotropic deformation potential and ionized impurities contribution within the Boltzmann transport theory, is obtained at a measured carrier density of $n_0 = n_{2D}d \sim$ 10$^{19}$ $\mathrm{cm}^{-3}$ (Fig. ~\ref{figS8}). The experimental $\Delta \mathrm{G}_\mathrm{BS}$ is approximated by the DC values as $\Delta \mathrm{G}_\mathrm{BS} \sim$ 1/$\Delta \gamma_\mathrm{BS}$, by fitting $\Delta \gamma_\mathrm{BS}$, since the contribution of $\Delta {n}$ is negligible, at 3\%–5\% of $n_{2D}$ (Fig.~\ref{figS4}). The experimentally determined $\Delta \mathrm{G}_\mathrm{BS}$ values correlate with those from the DFT calculation. The BS conductance slightly increases to around 4\% under tensile strain due to phase-space filling by the reduced $\mathrm{E}_{g}$. For larger strain, the BS undergoes a significant increase in conductance, indicating a metallic BS. Following this, the BS conductance ($\mathrm{G}_\mathrm{BS}$) again decreases due to the reopening of $\mathrm{E}_{g}$ after the band parity inversion, signifying the topologically trivial state ($\mathrm{Z}_2$ = 0) above ~5.5\% strain. This critical strain is similar to that observed in a 6 QL slab (Fig.~\ref{figS9}). Furthermore, the experimentally achieved maximum strain of approximately 7\% corresponds to the trivial insulator with an $\mathrm{E}_{g}$ of $\sim$100 meV. With a 9\% strain, $\gamma_\mathrm{BS}$ is greatly suppressed with a sufficient $\mathrm{E}_{g}$ ($\sim$200 meV), though this is not experimentally shown due to optical damage. Importantly, the light-driven strain can suppress the surface-bulk coupling by a synergetic combination of expanded phase space of the HTI and shrinking the BS. This allows for the separation of charge transport in TSS and BS, which is essential for topological applications at room temperature. In addition, the strain-induced conductance dynamics indicate the potential to transiently convert the conducting edge with an insulating bulk (TI) into the opposite configuration (the insulating edge with conducting bulk), and even achieve a NI phase at ambient conditions with a moderate $\mathrm{F}_\mathrm{pump}$ and doping level.

Following the transition, the expanded lattice damps with coherent interlayer vibrations as demonstrated in Figs.~\ref{fig1}b, ~\ref{fig1}d, and~\ref{fig3}a. The oscillation amplitude in $\Delta$R/$\mathrm{R}_0$ monotonically increases with an increasing, while the oscillations in $- \Delta$E/$\mathrm{E}_0$ become visible only at high $\mathrm{F}_\mathrm{pump}$. Figure~\ref{fig3}b presents the spectra of the measured interlayer vibrations, which are consistent with the interface mode (IM, 20 GHz) and breathing modes (BMs, 70 and 100 GHz)~\cite{21,22,23}. Despite this, only the IM drives the coherent modulation in THz conductance changes. The longitudinal strain waves are described by $\eta_{33}(\omega) = \eta_{33,0} \omega \delta /(1 + \omega^2\delta^2)$, where $\delta = v_s/\xi$. Here, $v_s$ = 2.4 nm/ps represents the longitudinal sound velocity and $\xi$ is approximately 20 nm~\cite{23}. The spectrum of strain waves encompasses the observed frequencies of interlayer vibrational modes, denoted with vertical lines. This indicates that a mode conversion efficiency between light-driven strain waves and eigenmodes in QL chains is higher for IM than BM. The $\mathrm{F}_\mathrm{pump}$-dependent amplitude of $\Delta \mathrm{R}$ ($\mathrm{A}_{\Delta \mathrm{R}}$), derived from Fig.~\ref{fig1}b, exhibits a linear relationship (Fig.~\ref{fig3}c). This is attributed to the proportional relationship between the $\mathrm{F}_\mathrm{pump}$ and $\eta_{33}$,0 (Fig.~\ref{figS6}). In terms of the $\mathrm{G}_0$ modulation (Fig.~\ref{fig3}d), $\mathrm{A}_{\Delta \mathrm{E}}$ also shows a linear increase above a threshold pump fluence ($\mathrm{F}_\mathrm{thres} \sim$ 1.4 mJ/$\mathrm{cm}^2$). It is noteworthy that the observed $\mathrm{F}_\mathrm{thres}$ corresponds with the $\mathrm{F}_\mathrm{pump}$ for the HTI transition (Fig.~\ref{fig2}d ). This suggests that the coherent modulation of transport can be realized near the transition point, where the conductance changes are sensitive to layer displacement.

\begin{figure*}[t]%
\centering
\includegraphics[width=0.9\textwidth]{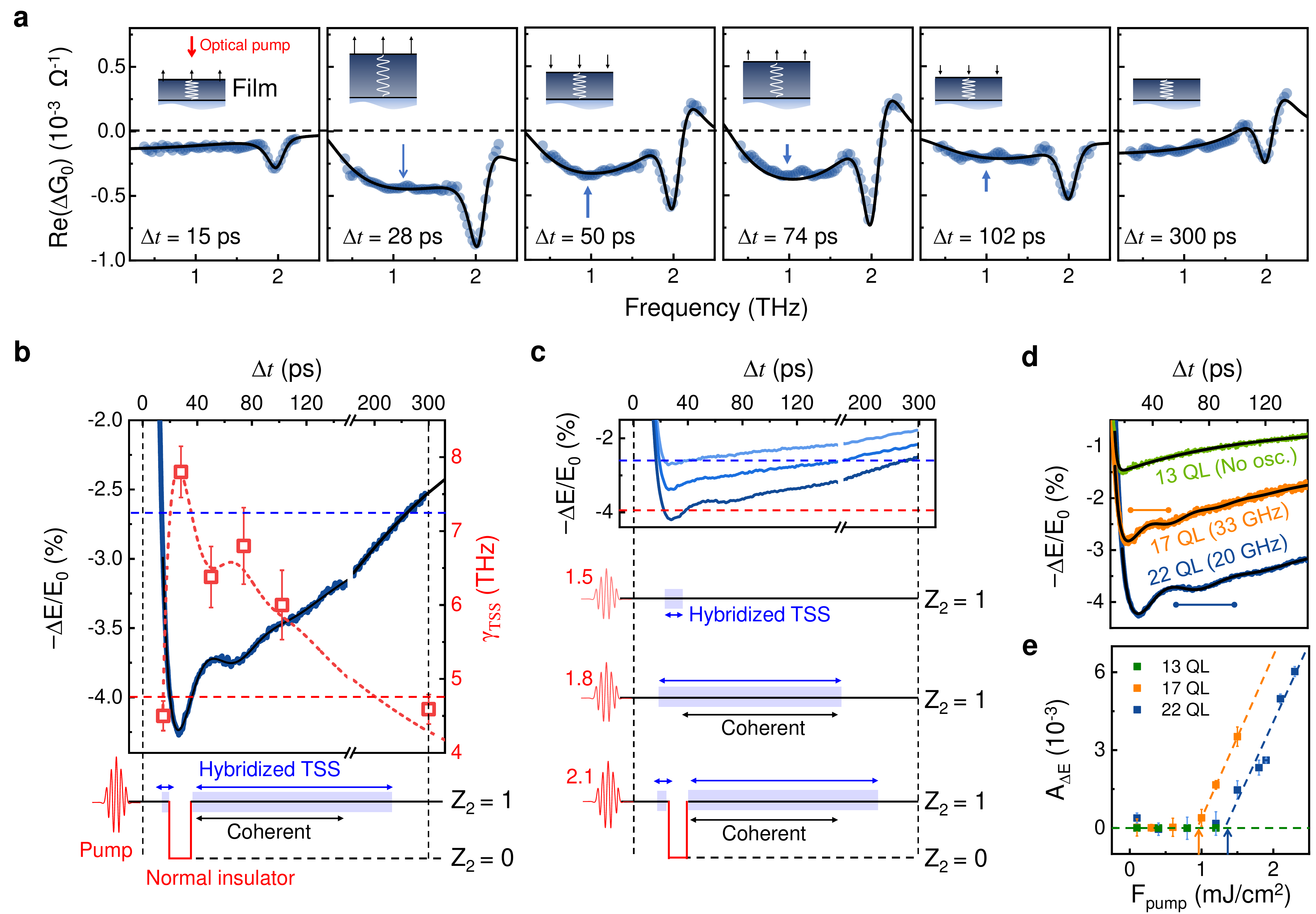}
\caption{
\textbf{Temporal sequences of topological-switching operation and interlayer vibration-assisted coherent modulation of TSS and BS transport.}
\textbf{a,} Time-dependent differential THz conductance at $\mathrm{F}_\mathrm{pump}$ = 2.3 mJ/$\mathrm{cm}^2$. $\Delta t$ is selected as during the expansion ($\Delta t$ = 15 ps), half period of IM vibrations ($\Delta t$ = 28 ps, 50 ps, 76 ps), and full relaxation ($\Delta t$ = 300 ps). The black curves are the Drude fits. 
\textbf{b,} Coherent modulation of THz conductance and the obtained scattering rate in TSS from (\textbf{a}). The error bars represent the standard deviation uncertainties of the fitting. The black curve is the fit result with an exponential decay with a time constant of $\tau_\mathrm{damping}\sim$ 300 ps and damped oscillations with IM frequency. The initial expansion and compression give $\sim$27\% modulation in transport lifetime in the TSS channel. The horizontal blue and red lines correspond to the scattering rates at HIT and NI phases adopted from Fig.~\ref{fig2}c inferred the relation between $- \Delta$E/$\mathrm{E}_0$ value and $\gamma_\mathrm{TSS}$. The bottom panel shows the temporal sequence of topological state dynamics. 
\textbf{c,} $\mathrm{F}_\mathrm{pump}$-dependent $- \Delta$E/$\mathrm{E}_0$ signal and time-dependent sequence of phases obtained in the same way from (\textbf{b}). 
\textbf{d} and \textbf{e}, Number of QL-dependent modulation frequencies (\textbf{d}) and amplitudes (\textbf{e}) measured in photoconductance. The modulation with 33 GHz frequency in 17 QL can be achieved, whereas such modulation is not obtained in 13 QL. The reduction of Fthres in 17 QL compared to that of 22 QL as well as the absence of modulation in 13 QL contain the finite size effects and the mode conversion efficiency of photoinduced strain waves to QL-dependent eigenmodes in linear chain model as discussed in the main text. The error bars in (\textbf{d}) and (\textbf{e}) represent the standard deviation uncertainties of the measured data.
}
\label{fig4}
\end{figure*}

This coherent modulation in transport is illustrated in Fig.~\ref{fig4}a. Here, the $\Delta \mathrm{G}_0$ spectra with $\mathrm{F}_\mathrm{pump}$ = 2.3 mJ/$\mathrm{cm}^2$ is displayed at the half period of the oscillation. After photoexcitation and carrier relaxation at $\Delta t$ = 15 ps, $\Delta \mathrm{G}_0$ exhibits a minor decrease during strain development. From $\Delta t$ = 28 ps (maximum expansion), $\Delta \mathrm{G}_0$ spectra manifest coherent oscillations that correspond to the lattice dynamics. Following $\Delta t =$ 102 ps, the coherence dissipates and steadily recovers toward equilibrium with damping, as portrayed in Fig.~\ref{fig4}b. The obtained $\gamma_\mathrm{TSS}$ dynamic suggests that the initial displacement vector of expansion transitions the topological phase toward the trivial insulator ($\mathrm{Z}_2$ = 0). The subsequent displacement, acting as a restorative force, restores it to the nontrivial state ($\mathrm{Z}_2$ = 1). The time-dependent $\gamma_\mathrm{TSS}$ displays coherent modulation while preserving the topological phase, characterized by the hybridized TSS, as depicted in the bottom panel of Fig.~\ref{fig4}b. Since the $n_{ex}$ after 10 ps is negligible, the $- \Delta$E/$\mathrm{E}_0$ signal is dominated by $\Delta \gamma_\mathrm{TSS}$ . This allows us to estimate the topological states based on $-\Delta$E/$\mathrm{E}_0$ values derived from Fig.~\ref{fig2}d. As a result, various time-dependent sequences of surface conduction and $\mathrm{Z}_2$ invariants after photoexcitation with several $\mathrm{F}_\mathrm{pump}$ values can be obtained, as shown in Fig.~\ref{fig4}c. For $\mathrm{F}_\mathrm{pump}$ = 1.5 mJ/$\mathrm{cm}^2$, the strain approaches the hybridization point before transition, generating a transient topological state with the hybridized TSS for around 15 ps. This duration can be extended beyond 100 ps in conjunction with the occurrence of coherent modulation at a higher $\mathrm{F}_\mathrm{pump}$ of 1.8 mJ/$\mathrm{cm}^2$. In the case of $\mathrm{F}_\mathrm{pump}$ = 2.1 mJ/$\mathrm{cm}^2$, where the temporal sequence is comparable to $\mathrm{F}_\mathrm{pump}$ = 2.3 mJ/$\mathrm{cm}^2$ as seen in Fig.~\ref{fig4}b, the topological phase transitions to a topologically trivial state for approximately 10 ps. This is followed by the hybridized TSS for $\sim$100 ps, coinciding with the coherent modulation of transport properties. Thus, by merely adjusting $\mathrm{F}_\mathrm{pump}$ as a control parameter, we are able to achieve various temporal sequences of the transient HTI or $\mathrm{Z}_2$ switching.

The switching operation observed can be attributed to the mode conversion (from light to strain waves) and the finite size effect. Thinner films display blue-shifted eigenmodes of interlayer vibrations~\cite{21,22,23}, enabling a faster modulation for the 17 QL Bi$_2$Se$_3$, as depicted in Fig.~\ref{fig4}d. However, such modulation was not observed in the 13 QL, potentially due to a poor mode conversion, because the eigenmode frequency for 13 QL is 50 GHz~\cite{28}, far from the peak of spectrum of the photoinduced strain. To achieve faster modulation with the 13 QL, pump wavelength tuning might be necessary to blue-shift the strain spectrum. Furthermore, the finite size effect suggests that hybridization occurs at a relatively larger $\mathrm{E}_{g}$ in thinner films, as shown in Fig.~\ref{fig4}e. The $\mathrm{F}_\mathrm{thres}$ for the 17 QL is smaller than that of the 22 QL, approximately 1.0 mJ/$\mathrm{cm}^2$, corresponding to a strain of around 3\%. For a 3\% tensile strain, $\mathrm{E}_{g}$ is 150 meV (Fig.~\ref{figS8}) and $\lambda_\mathrm{TSS}$ extends to roughly 8 nm (half of the 17 QL thickness). It is expected that the top and bottom TSSs have already partially overlapped for roughly 10 QL or less~\cite{6,39}, explaining the lack of coherent modulation in the 13 QL. Similarly, a bulk crystal might not be suitable for the switching operation because induced strain waves propagate without confinement and subsequent lattice expansion~\cite{25}. Therefore, the film thickness (above ~20 QL) and pump wavelength play crucial roles in this switching operation. This methodology can be further developed by combining with films at phase boundaries~\cite{6,7} and heterointerfaces~\cite{40,41}, and utilizing an ultrafast coherent control scheme~\cite{21,42} to create ultrafast topological switching devices for a variety of applications.

\begin{acknowledgments}
This work was supported by the National Research Foundation of Korea (NRF) funded by the Ministry of Science, ICT \& Future Planning, Korea (RS-2023-00208484, 2019M3D1A107830222, 2019R1A6A1A10073887, 2021R1F1A1050726, 2021R1A2C2004324) and the National Research Council of Science \& Technology (NST) grant from the Korean Government (CAP18054-000).

\end{acknowledgments}
\bibliography{TPT.bib}


\setcounter{figure}{0}
\renewcommand{\figurename}{Fig.}
\renewcommand{\thefigure}{S\arabic{figure}}

\newpage
\pagebreak 
\widetext
\clearpage

\Large{{Supplementary Material for}

\begin{center}
\large{\bf{Ultrafast switching of topological invariants by light-driven strain}}
\end{center}

\section{EXPERIMENTAL METHOD AND DATA ANALYSIS}
\subsection{Sample preparation}

\normalsize{
Bi$_2$Se$_3$ thin films were prepared employing the molecular beam epitaxy (MBE) technique. The formation of crystalline and thickness of Bi$_2$Se$_3$ were characterized by an in-situ reflections high-energy electron diffraction (RHEED) system and atomic force microscopy (n-Tracer, Nano Focus Inc.) in a non-contact mode. A detailed description of preparation and characterization can be found in our previous work~\cite{21}.
}

\subsection{Ultrafast optical and THz spectroscopy}

\normalsize{
Ultrafast optical pump and NIR probe measurements were conducted by using a mode-locked Ti:sapphire oscillator (MAITAI, Spectra-Physics), which generates 100-fs pulses at a central wavelength of 830 nm and 80 MHz repetition rate. A partial output was used for pumping a synchronously pumped optical parametric oscillator, which generated 150-fs probe pulses (1350 nm) at the same repetition rate. After collinearly combining by a dichroic mirror, the pump and probe pulses were focused on the sample to a beam waist of $\sim$2 $\mu$m by a single objective lens. The time-resolved optical pump and THz probe spectroscopy was performed by utilizing 100-fs laser pulses generated from a 1-kHz Ti:sapphire regenerative amplifier (Spitfire Ace Pro, Spectra Physics) operating at 800 nm. THz pulses serving as the probe beam were generated by optical rectification in a 2-mm-thick ZnTe and detected by electro-optic sampling. The differential reflection of NIR probe pulses was measured with a Ge photodetector (DET50B, Thorlabs), while THz signal was detected in transmission geometry by using a lock-in-Amplifier (SR830, Stanford Research Systems) with a modulation at 190 Hz. The time delay with the pump pulses was controlled by a motorized stage. All optical experiments were conducted under ambient conditions at room temperature and normal pressure, while THz experiments were performed under 2\% humidity by purging the entire THz beam path with dry air. For the analysis of THz signal, we applied thin-film limit ($\lambda_\mathrm{THz}\gg\ d_\mathrm{sample}$). Hence, THz complex conductance $\Tilde{G}_0(\omega)$ of sample was directly deduced from the complex transmission ($\Tilde{T}(\omega)=\mathrm{E}_\mathrm{sam+sub}(\omega)/\mathrm{E}_\mathrm{sub}(\omega)$) of THz field as given by~\cite{6,27} $\Tilde{G}_0(\omega)=\frac{1+n_{sub}}{Z_0 d_{sam}}(\frac{1}{\Tilde{T}(\omega))-1})$ where $n_{sub}$ is the refractive index of the substrate at THz frequency. $Z_0 \cong 377 \Omega$ is the vacuum impedance. $d_sam$ is the thickness of the sample. The spectrum of complex conductance in Bi$_2$Se$_3$ can be fitted by the Drude model, consisting of one Drude term from the topological surface state (TSS) and one Drude-Lorentz term from phonon resonance, given by 

\begin{center}
$\Tilde{G}_0(\omega)=\Bigr[-\frac{\omega_{pD}}{i\omega-\Gamma_D}-\frac{i\omega\omega_{pL}^2}{\omega_L^2-\omega-i\omega\Gamma_L}-i(\epsilon_{\infty-1})\Bigr] \epsilon_0 d$	
\end{center}
	
where $\omega_{pD}$ and $\Gamma_D$ are the plasma frequency and scattering rate (inverse of transport lifetime) for the Drude term. $\omega_{pL}$ and $\Gamma_L$ are the plasma frequency and scattering rate Drude-Lorentz term. $\epsilon_0$ is the vacuum permittivity and $d$ is the film thickness. The changes in conductance by photoexcitation was obtained with differential THz transmission, $\Delta\Tilde{t}(\omega)=[\Tilde{E}_{\mathrm{pump}}(\omega)-\Tilde{E}_{\mathrm{no} \mathrm{pump}}(\omega)$]. The conductivity changes can also be fitted with the Drude model, which gives information such as the dynamic changes of plasma frequency and scattering rate of Bi$_2$Se$_3$ film by photoexcitation and their recovery. The changes of conductance ($\Delta\Tilde{G}_0$) in Bi$_2$Se$_3$ can also be analyzed by fitting with the Drude-Lorentz model as ($\Delta\Tilde{G}(\omega,t)=\Tilde{G}(\omega,t)-\Tilde{G}_0(\omega)$) where, $\Tilde{G}(\omega,t)$ and $\Tilde{G}_0(\omega)$ are THz conductance with and without optical pump. Drude-Lorentzian fit for measured spectra of photoconductance was conducted by using the RefFIT program~\cite{43}.
}

\subsection{Density functional theory (DFT) calculation}
\normalsize{
The Vienna Ab initio Simulation Program (VASP) was used for the density functional theory calculation with the Perdew-Burke-Ernzerhof (PBE) functional~\cite{44,45}. We used the plane-wave energy cutoff of 600 eV for total energy and band structure calculations of Bi$_2$Se$_3$ bulk with the van der Waals (vdW) interaction corrected by the DFT-D3 method and spin-orbit coupling~\cite{46}. The $\Gamma$-centered 20 × 20 × 1 Brillouin zone sampling was used. The total energy and ionic force relaxations were performed under the criteria of $10^{-8}$ meV and $10^{-4}$ eV/\r{A}, respectively. To demonstrate the strain-dependent phase transition of surface states (Fig.~\ref{figS9}), we enforce a strain of 6 QL by fixing atomic positions in a slab model. To calculate electrical conductivity, we used Ab initio Scattering and Transport (AMSET) software with fully anisotropic consideration of acoustic deformation poten-tial and ionized impurities scatterings based on Boltzmann transport equation~\cite{47}.
}

\section{Photoinduced strain waves and interlayer vibration effects}
\normalsize{
As discussed in the main text, only longitudinal strain waves ($\eta_{L} = \eta_{33}$) are considered in our experiments. The generation mechanisms of photoinduced stress by ultrafast laser are based on thermoelasticity (TE), deformation potential (DP), and inverse piezo effects~\cite{37}. The inverse piezo effects dominate in non-centrosymmetric crystals and are irrelevant in Bi$_2$Se$_3$ crystals. From the microscopic point of view, the photoinduced stress can be formed in all directions from a photoexcited point. As a result, the atomic displacement will be perturbed with quasi-spherical symmetry due to laterally isotropic surface of Bi$_2$Se$_3$. Then, the shear strain ($\eta_{S} = \eta_{11}$ and $\eta_{22}$) in Fig.~\ref{figS1}a cannot be excited because the transverse displacement is orthogonal to the spherical symmetry, making itself negligible with the condition of $A_\mathrm{pump} \gg \xi_\mathrm{pump}$, where $\xi_\mathrm{pump}$ is the optical penetration depth at the pump wavelength and $A_\mathrm{pump}$ is the excitation area of the pump beam. For this reason, we treated the photoinduced strain wave as a “plane” wave along the longitudinal propagation direction as negligible quasi-shear strain ($\eta_{L} = \eta_{13,31}$ and $\eta_{23,32}$) shown in Fig.~\ref{figS1}b. A similar physical phenomenon, satisfying the spherical symmetry, can be seen in wave diffraction at a slit, which is explained by Huygens’s principle. Accordingly, a wave passing through a broad (or narrow) slit proceeds as a plane (or spherical) wave. The propagation direction of the strain wave will be dominant along the longitudinal direction in the case of $A_\mathrm{pump} \gg \xi_\mathrm{pump}$ and normal incidence of pump pulses. Also, $\eta_{33}$ confined inside the film forms standing waves and acts as an effective selective tensile strain when the film thickness ($d$) is reduced. Consequently, $\eta_{33}$ is mainly attributed to the topological phase transition as well as the initiation of interlayer vibrational modes. As known, Bi$_2$Se$_3$ contains the layered rhombohedral crystal structure with $a=b\neq c$, $\alpha=\beta=90°$, and $\gamma=120°$. The interlayer shear mode would be further generated by breaking the symmetry of in-plane geometry. Thus, only out-of-plane strain and vibrations have been also observed in this work as reported in the previous ultrafast optical~\cite{21} and X-ray studies~\cite{22}.

\begin{figure*}[t]%
\centering
\includegraphics[width=0.5\textwidth]{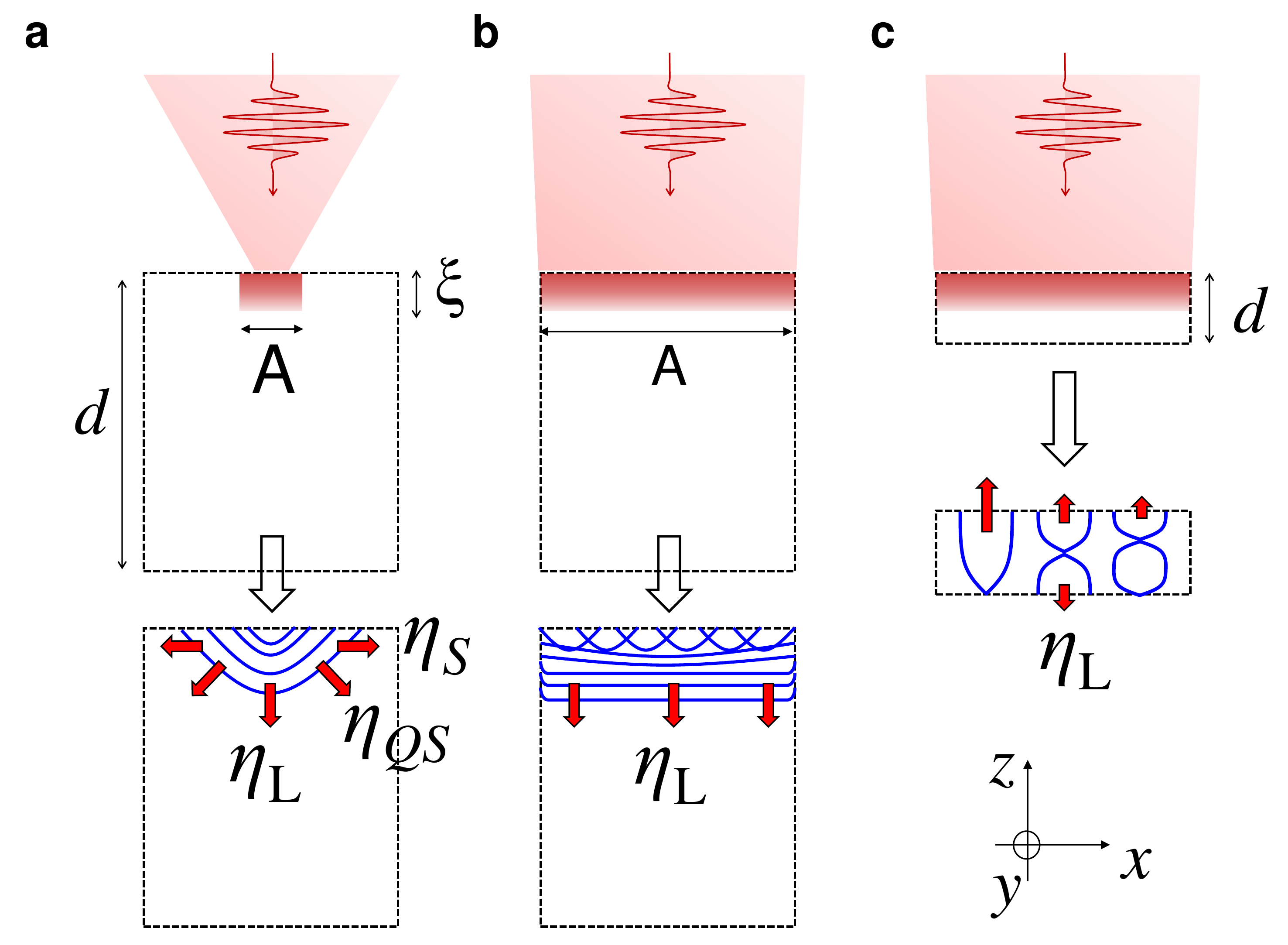}
\caption{
\textbf{Schematic of strain waves generated by light excitation in case of}
\textbf{a,} $\xi_\mathrm{pump} \sim \mathrm{A}_\mathrm{pump}$ and \textbf{b,} $\xi_\mathrm{pump} \ll \mathrm{A}_\mathrm{pump}$, and \textbf{c,} confinement of strain waves by lack of translation symmetry in nanoscale thickness.
}
\label{figS1}
\end{figure*}

Under tensile strain in Bi$_2$Se$_3$, both QL thickness and inter-QL distance change. From density functional calculations, the relaxed QL thickness is 7.06 \r{A} and the inter-QL distance (van der Waals gap) is 3.10 \r{A}. As shown in Fig.~\ref{figS2}, the DFT calculation revealed that the inter-QL distance changes up to ~20\% under 5\% tensile strain, while the QL thickness is almost unaltered. Hence, the modulation in the inter-QL distance is responsible for the transition between topologically nontrivial and trivial phases. The tensile strain changes the electronic band structure as shown in Fig.~\ref{fig2}e in the main text. Subsequent optical constants change the complex refractive index ($\Tilde{n}=n+i\kappa$) or the dielectric constant ($\epsilon=\Tilde{n}^2$) and become $\Delta n(z,t) = \frac{\partial n}{\partial \eta_{33}} \eta_{33}(z,t)$ and $\Delta\kappa(z,t) = \frac{\partial \kappa}{\partial \eta_{33}} \eta_{33}(z,t)$. In the detection scheme, optical transmission or reflection is measured. From that, we are able to estimate the transmission and reflection coefficient as $t_0 = 2/(1+n+i\kappa)$ and $r_0 =(1+n-i\kappa)/(1+n+i\kappa)$ based on the Fresnel equation. In this work, we measured coherent phonons using reflection coefficient changes by strain waves in a reflection geometry, which can be expressed by~\cite{48} $$\Delta r/r_0 \propto \frac{d\Tilde{n}}{d\eta_{33}} \int_{0}^{d} \eta_{33} exp(2i\Tilde{n}k_{pr}z) \,dz$$ where $k_{pr}$ is the wave vector of probe pulses and $d$ is the sample thickness. The real part is a measurable quantity in optical experiments by interference with the reflected light at the sample surface. The amplitude of the observed coherent reflection modulation is proportional to the term of $d\Tilde{n}/d\eta_{33}$, namely, the photoelastic coefficient. The measured oscillation amplitude ($\mathrm{A}_\mathrm{osc}$) linearly depends on $d\Tilde{n}/d\eta_{33}$. Subsequently, refractive index changes are determined by bandgap modulation ($\delta E_{g}$). Hence, the differential optical reflection ($\Delta$R/$\mathrm{R}_0$) records the changes in a real part of the refractive index ($\Tilde{n}$) by deformation. The signal will be strong in the vicinity of the bandgap (i.e. $\mathrm{A}_\mathrm{osc}\sim d\tilde{n}/d\omega\cdot\delta E_{g})$)~\cite{26} as shown in Fig.~\ref{figS3}.

\begin{figure*}[t]%
\centering
\includegraphics[width=0.6\textwidth]{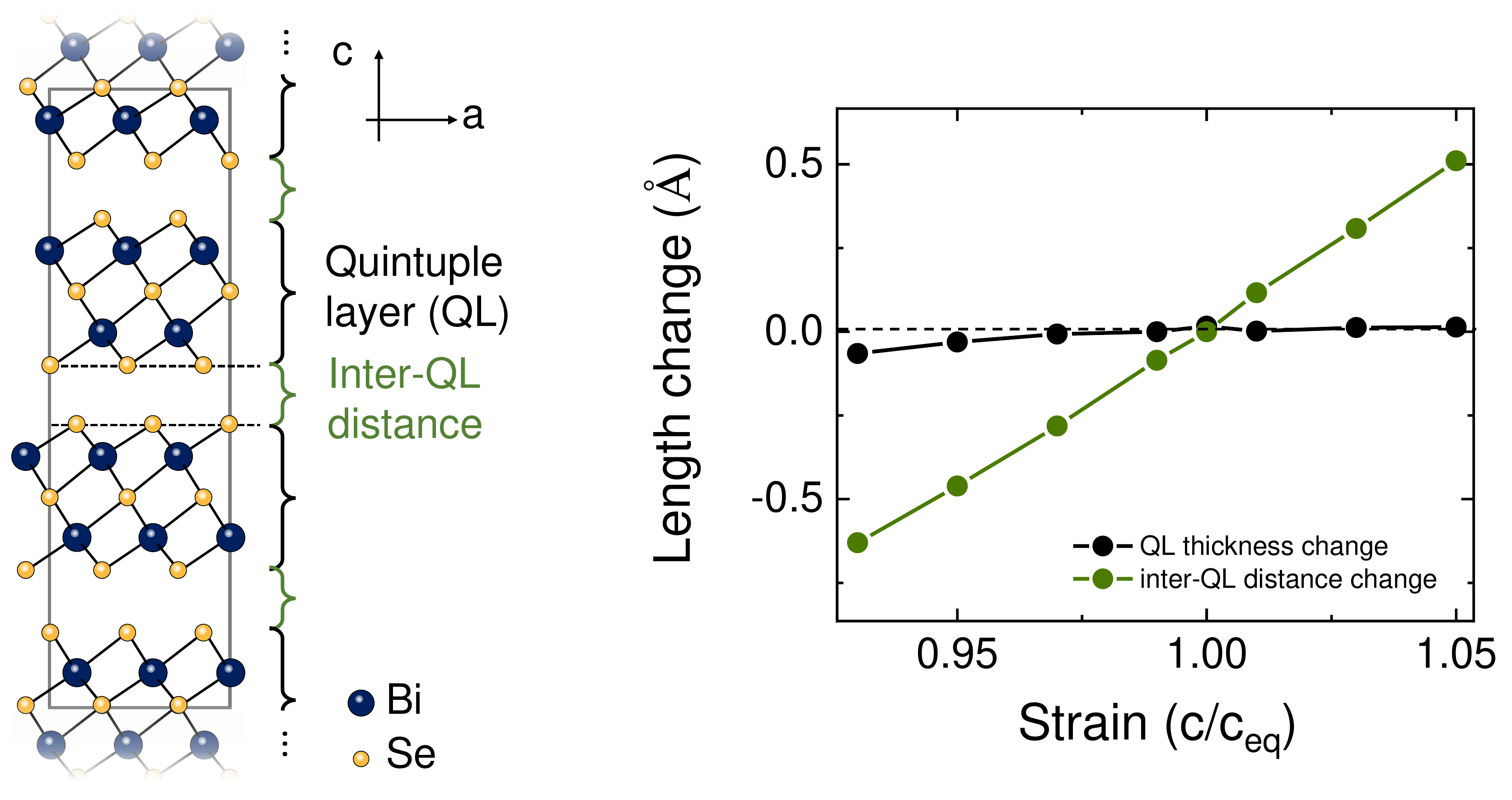}
\caption{
\textbf{Crystal structure of Bi$_\textbf{2}$Se$_\textbf{3}$ consisting of quintuple layer as one unit with periodic van der Waals gap as inter-QL distance.}
Under out-of-plane tensile strain, the inter-QL distance sharply changes compared to each QL thickness as a factor of ~40.
}
\label{figS2}
\end{figure*}

 The electronic stress can be formed by electron-lattice interaction. In addition, the generated photoinduced stress propagates inside the film with a finite sound velocity ($v_s$). Thus, an actual expansion of lattice by photoinduced strain waves requires proper time after photoexcitation. In 22 QL Bi$_2$Se$_3$, a maximum expansion is observed at $t_{2}$. This tensile straining in film induces an increase in scattering rate. The Drude-Lorentzian fit gives $\gamma_\mathrm{TSS,BS}$ values and $D_\mathrm{TSS,BS}$ at $t_{2}$ (expansion). Since the carrier relaxation is extremely fast, the remaining excited carriers can be expected to be negligibly small. Figure~\ref{figS4} shows the density of unrelaxed carriers at $t_{2}$. In the case of TSS, the unrelaxed carriers show a linear dependence on $\mathrm{F}_\mathrm{pump}$. This linear increase confirmed that the carrier relaxation time is constant as a function of the carrier density. The bulk carrier is relaxed by optical phonon scattering. Hence, the relaxation time depends on electron-phonon coupling and phonon energy, which would be irrelevant to excited carrier density. The slope of the linear fit in $D_\mathrm{TSS}$ indicates the unrelaxed carrier density at $t_{2}$ kept at ~0.12\% of $n_ex$. This means that the photocarrier is mostly relaxed. In addition, $D_\mathrm{BS}$ is also captured by straining. The positive of $\Delta \mathrm{G}_0$ in Fig.~\ref{fig2}a of the main text is reproduced by decreased $\gamma_\mathrm{BS}$ with small $D_\mathrm{BS}$ (~2-5\% of n0) as shown in Fig.~\ref{figS4}. The background carrier density in BS cannot be estimated, since the carrier density in BS and TSS degenerates in the THz signal at room temperature by surface-bulk coupling. However, since the measured Fermi energy is slightly above $\mathrm{E}_{g}$, it would also be expected that the bulk carriers at equilibrium are negligible compared to the background carriers at the surface. This straining reduces the THz photoconductance as shown in Fig.~\ref{figS4}a. Thus, a decay time of reduced photoconductance represents the damping time of strain waves in films. The obtained damping time of ~300 ps is consistent with the previous X-ray results~\cite{22}.
}

\begin{figure*}[t]%
\centering
\includegraphics[width=0.6\textwidth]{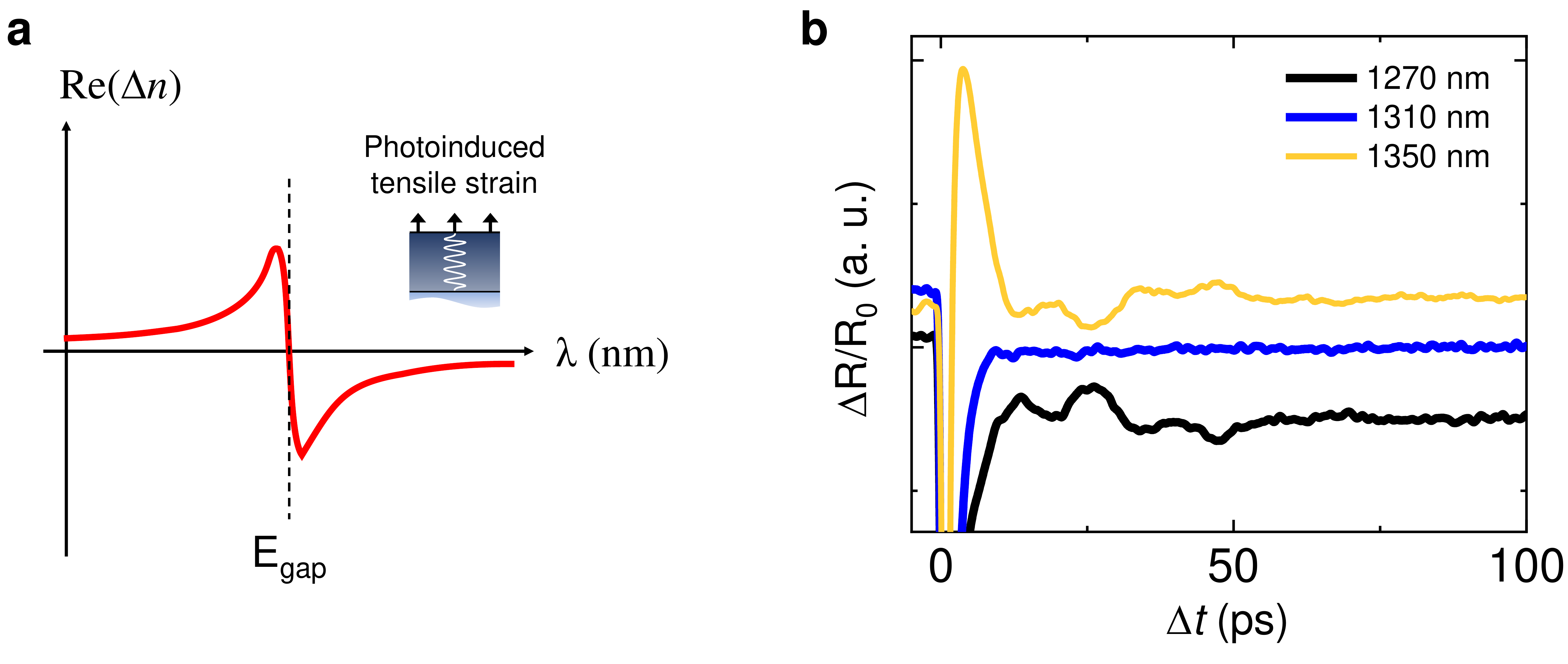}
\caption{
\textbf{a}, Measured $\Delta$R/$\mathrm{R}_0$ with different probe wavelengths \textbf{b}, Schematic of real part of modified refractive index near bandgap. Coherent vibration signals with opposite phases (at 1270 and 1350 nm) or zero intensity (at 1310 nm) can be observed in the same strain according to probe wavelength.
}
\label{figS3}
\end{figure*}

\begin{figure*}[t]%
\centering
\includegraphics[width=0.5\textwidth]{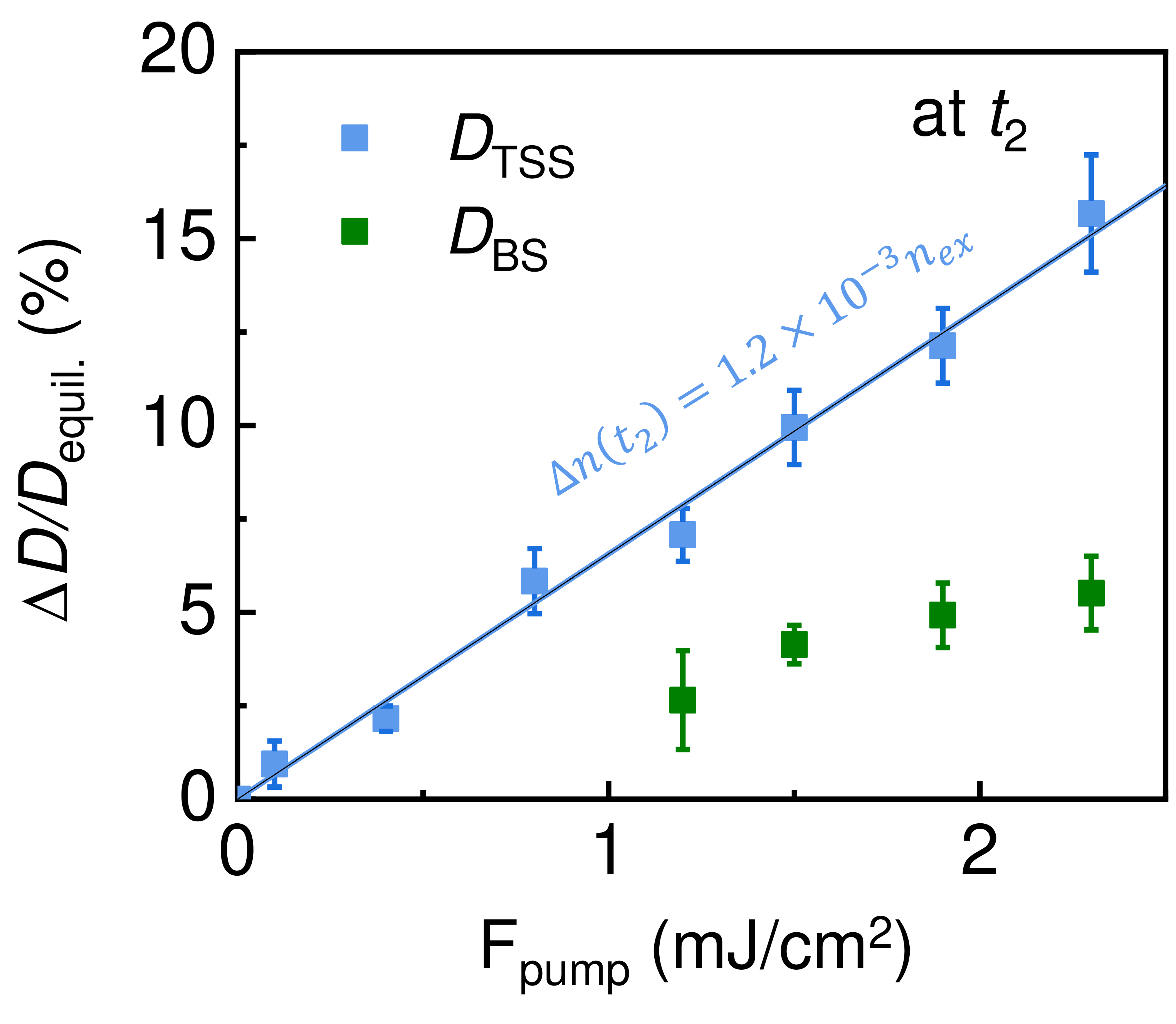}
\caption{
$\mathrm{F}_\mathrm{pump}$-dependent $\Delta D_\mathrm{TSS}$, BS with a linear fit. $D_0$ = 138 THz$^2$$\cdot$QL corresponds to $n_0$ = 1.78×10$^{13}$ cm$^{-2}$ at equilibrium. $\Delta D$ is from excited carrier density ($n_{ex}$ = 9.14×10$^{14}$ cm$^{-2}$ for 1 mJ/cm${^2}$ photoexcitation). Hence, the slope indicates that the unrelaxed carrier population at $t_{2}$ is ~0.12\% of $n_{ex}$.
}
\label{figS4}
\end{figure*}

\begin{figure*}[t]%
\centering
\includegraphics[width=0.9\textwidth]{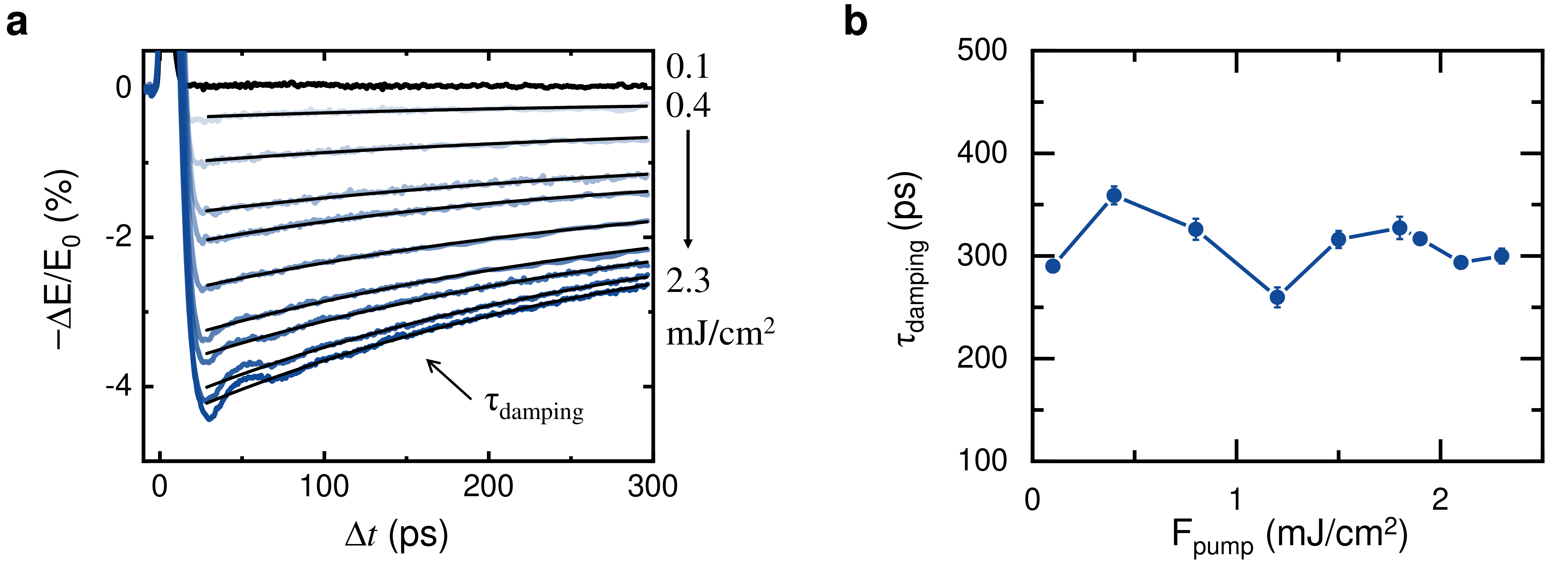}
\caption{
\textbf{a},Differential THz signal empathizing the decreased conductance at different $\mathrm{F}_\mathrm{pump}$. 
\textbf{b}, $\mathrm{F}_\mathrm{pump}$-dependent damping time obtained by fitting with exponential decay from (\textbf{a}).
}
\label{figS5}
\end{figure*}

\section{Calculation of photoinduced stress and strain amplitudes}
\normalsize{
Generation mechanisms of photoinduced strain ($\eta =dV/V_0$, where $V$ is the volume) by laser illumination consider electronic stress, lattice stress, and inverse piezo effects~\cite{37}. The inverse piezo effect is not relevant to the crystal structure of Bi$_2$Se$_3$. In the case of the electronic stress by deformation potential mechanism, sudden changes in electronic distribution by light absorption induce electronic stress. The excited electrons and holes can modify the neighboring atomic interaction from the equilibrium. This modification also changes atomic distance and leads to crystal deformation ($\sim dE/d\eta$). Under photoexcitation with a 1.5-eV photon energy above the Bi$_2$Se$_3$ bandgap ($E_{g}\sim$~0.3 eV), the excited electrons in the valence band move into the conduction band. This distribution changes from the excited carriers ($N_{ex}$) lead to the electronic stress ($\sigma_{DP}$) and are simplified with a deformation potential ($D$) as $$\sigma_{DP} = \sum_{k} \delta n_{ex} (k) \frac{d E_{k}}{d\eta} = -Dn_{ex}$$ Here, we ignore electronic distribution modification by electron temperature ($T_e$), which is determined by Fermi-Dirac distribution with Te changes and expressed with the electron Grüneisen coefficient ($\gamma_e$) by $\sigma_{DP}=-\gamma_{e} C_{e}\Delta T_{e}$. Because the thermalization time of electrons by electron-electron scattering is typically within a few tens of femtoseconds, it is too fast to efficiently generate acoustic stress. Hence, we consider the excited carrier density ($n_{ex}$) to calculate $\sigma_{DP}$ in 22 QL Bi$_2$Se$_3$. For the optical pump pulses, the excited carrier density ($n_{ex}$) along the z-direction can be obtained from the continuity equation as $P=-\bigtriangledown_{z} I$ and $I = I_0 (1 - R) exp(-z/\xi_{pump}$), where $R$ is the reflectivity ($R \sim$ 0.4) and $\xi_{pump}$ is the optical penetration depth from absorption coefficient ($\alpha \sim$ 5×10$^5$ $\mathrm{cm}^{-1}$) at pump wavelength (800 nm)~\cite{49}. The DP stress is directly achievable by multiplying $n_{ex}$ with the deformation potential ($D$ = 22 eV)~\cite{50}.

Lattice stress is based on the thermal expansion of matter by thermoelasticity (TE). Under photoexcitation, the excited “hot” electrons will relax their energy by interacting with the environment (mainly lattice). In other words, the temperature of the electronic subsystem decreases by electron-phonon scattering, and then, the lattice temperature increases ($\Delta T_{L}$). This leads to thermal expansion and related thermoelastic stress ($\sigma_{TE}$), which is given by $\sigma_{TE}=-3 \beta B \Delta T_L$ where $\beta$ and $B$ are the linear expansion coefficient and bulk modulus with a unit of $K^{-1}$ and GPa, respectively. In addition, the lattice temperature can be calculated by the temperature model according to the coupled partial differential equations of electron temperature ($T_e$) and lattice temperature ($T_L$) with heat capacity for electron and lattice ($C_e$ and $C_p$). In the case of $\sigma_TE$, the changes in lattice temperature ($\Delta T_L$) can be estimated from the relation of $\Delta T_L \sim \mathrm{F}_\mathrm{pump}/(\xi C_p)$~\cite{51}. Here, we used the linear expansion coefficient ($\beta$) of 4.1×$10^{-5}$ K$^{-1}$ and the heat capacity ($C_p$) of 2.8 JK$^{-1}$cm$^{-3}$ at room temperature taken from Ref. 
~\cite{52,53}. Note that the obtained $\sigma_t$ was overestimated because the heat capacity becomes larger as the lattice temperature increases ($C_p \propto T_L$ at high temperature). 
}

\begin{figure*}[t]%
\centering
\includegraphics[width=0.8\textwidth]{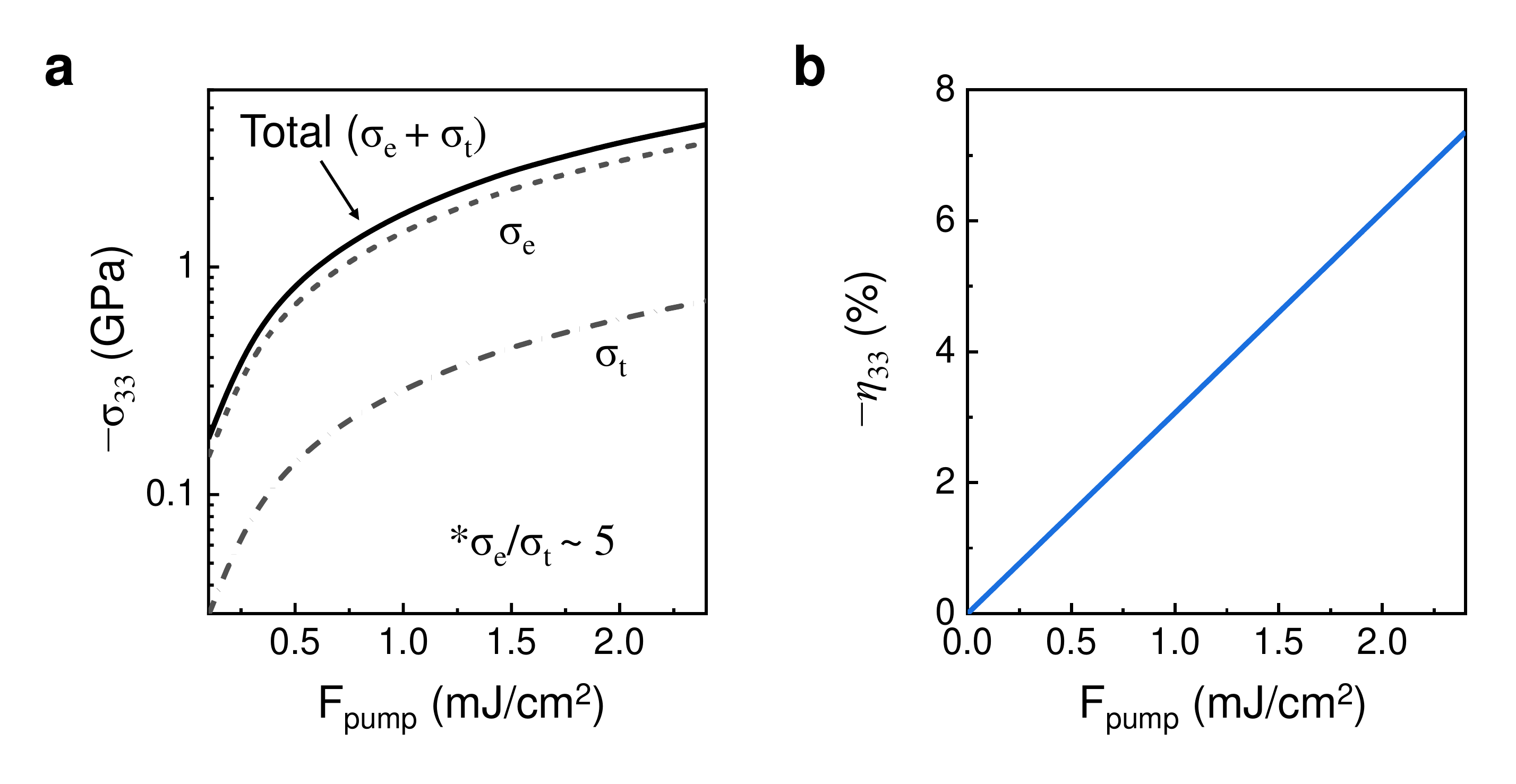}
\caption{
\textbf{a}, Calculated photoinduced stress as a function of $\mathrm{F}_\mathrm{pump}$ based on deformation potential (DP) and photoelasticity (TE), which generates electronic ($\sigma_e$) and thermal stress ($\sigma_t$) in film. The ratio $\sigma_e/\sigma_t \sim$ 5 indicates that the electronic stress dominates in photoinduced stress, which leads to a linear relationship between $\sigma_{33}$ and $\mathrm{F}_\mathrm{pump}$ ($\sigma_{e} \propto n_{ex} \propto \mathrm{F}_\mathrm{pump}$).  
\textbf{b}, Corresponding amplitude of strain obtained by dividing (\textbf{a}) with the bulk modulus ($C_{33}$).
}
\label{figS6}
\end{figure*}

\begin{figure*}[t]%
\centering
\includegraphics[width=0.7\textwidth]{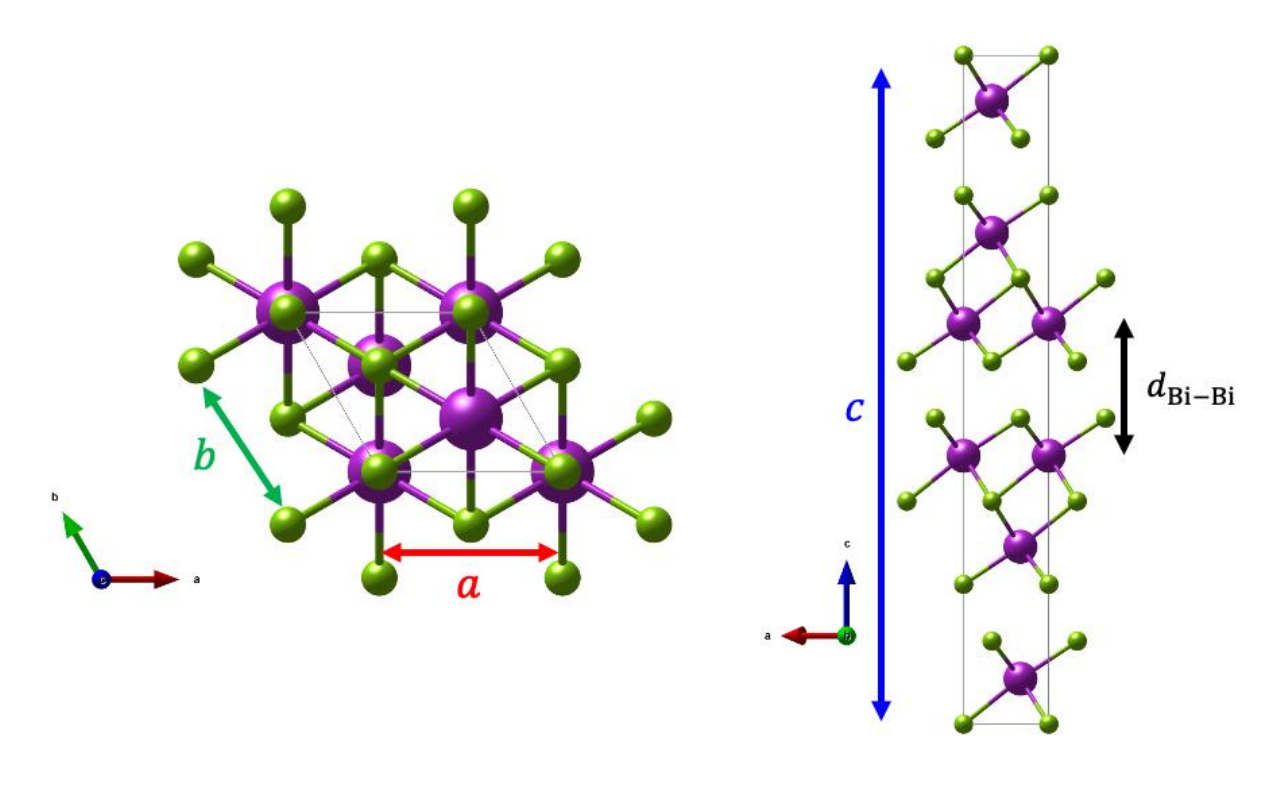}
\caption{
Atomic structure for electrical conductivity calculation drawn using VESTA~\cite{54}. The in-plane lattice constants are relaxed for a fixed interlayer distance. The relative angle between a and b lattices is 120°. Detailed information on atomic structures for different strains is listed in Table S1.
}
\label{figS7}
\end{figure*}

Table \ref{demo-table} has a caption:
\begin{table}[!h]
\begin{center}
\begin{tabular}{||c c c c||}

 \hline
 Strain(\%) & $a=b (\r{A})$ & $c (\r{A})$ & $d_{Bi-Bi}$ \\ [0.5ex] 
 \hline\hline
 0 & 4.153 & 28.225 & 5.570 \\ 
 \hline
 1.5 & 4.232 & 28.667 & 5.724 \\
 \hline
 3.3 & 4.220 & 29.171 & 5.867 \\
 \hline
 5.1 & 4.211 & 29.676 & 6.019 \\
 \hline
 7.2 & 4.203 & 30.248 & 6.193 \\
 \hline
 9.0 & 4.198 & 30.753 & 6.353 \\ [0.5ex] 
\hline

\end{tabular}
\caption{\label{demo-table}The lattice constants for different strained systems}
\end{center}
\end{table}

\begin{figure*}[t]%
\centering
\includegraphics[width=0.8\textwidth]{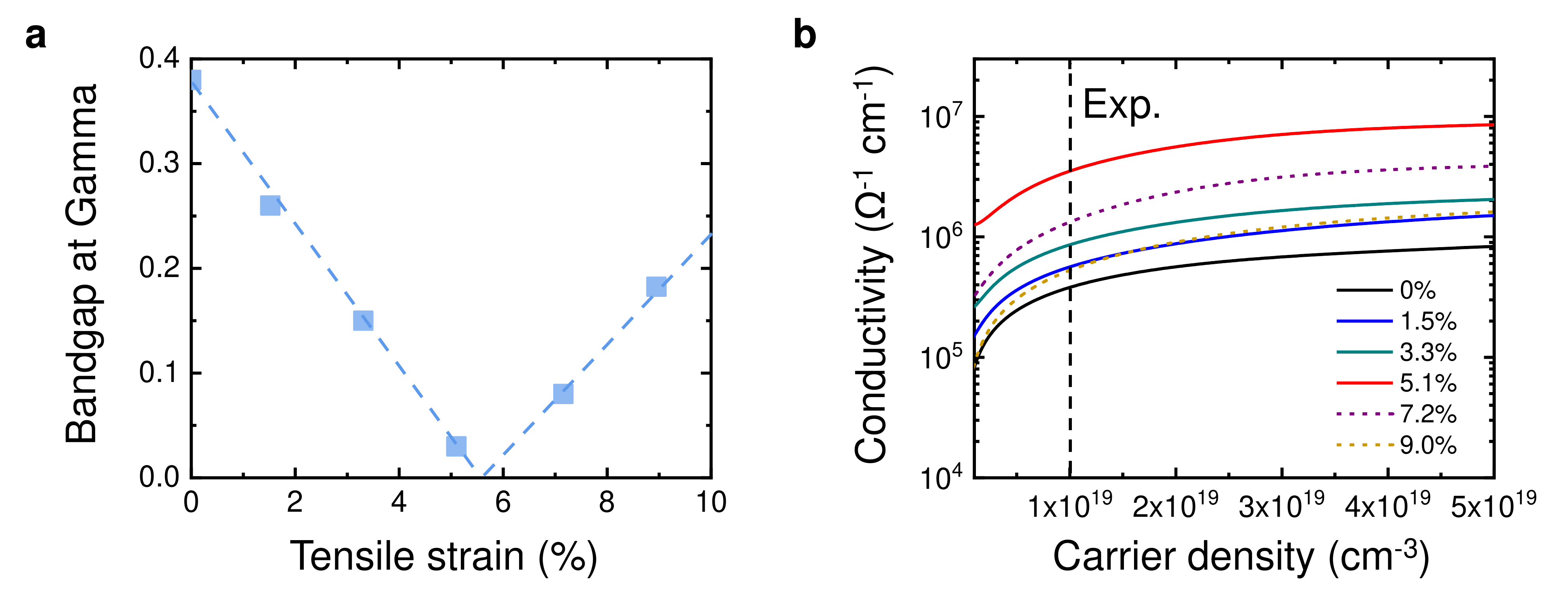}
\caption{
\textbf{a}, Bandgap at Gamma point. 
\textbf{b}, Bulk conductivity according to tensile strain obtained from calculated band structure in Fig.~\ref{2}b in the main text.
}
\label{figS8}
\end{figure*}

\begin{figure*}[t]%
\centering
\includegraphics[width=0.9\textwidth]{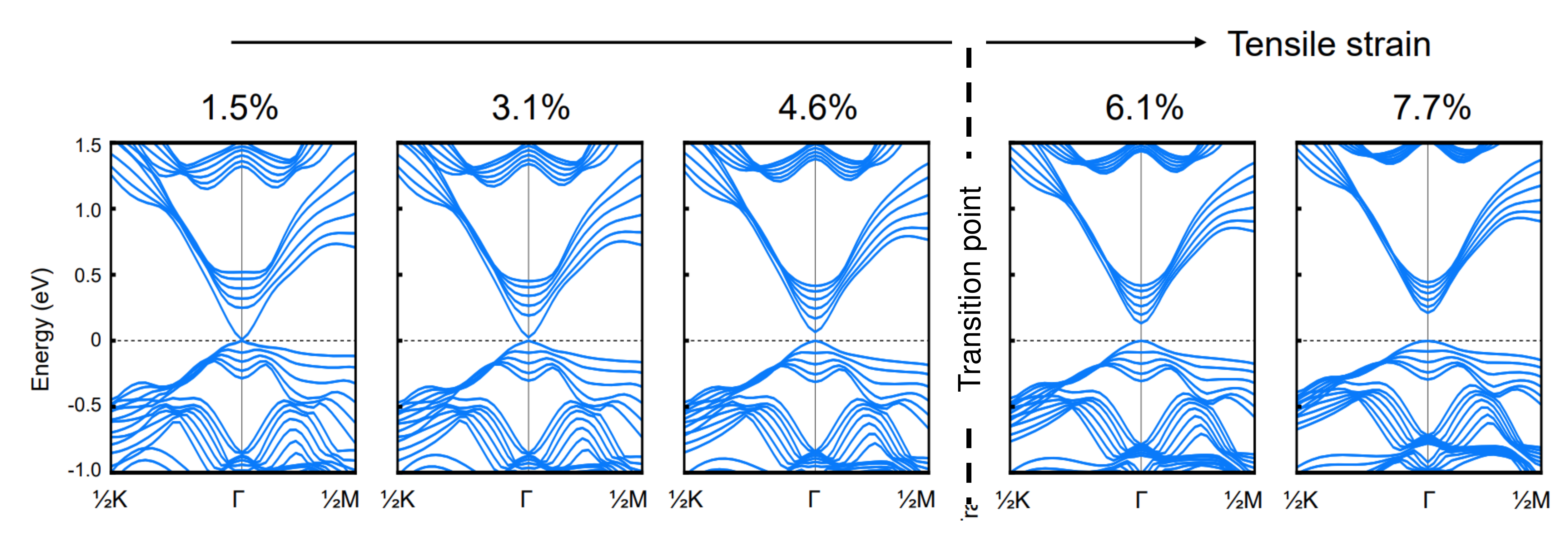}
\caption{
Calculated strain-dependent band structure in 6 QL slab.
}
\label{figS9}
\end{figure*}

\end{document}